\documentclass[aps,prb,twocolumn,superscriptaddress]{revtex4-1}

\newcommand{\figurescale}{1}

\usepackage{verbatim}
\usepackage{amsmath}
\usepackage[utf8]{inputenc}
\usepackage[pdftex]{graphicx}
\usepackage[separate-uncertainty=true]{siunitx}
\DeclareSIUnit{\rpm}{rpm}
\usepackage[colorlinks=true,urlcolor=blue,linkcolor=blue,citecolor=blue]{hyperref}
\usepackage{xcolor}
\usepackage[T1]{fontenc}
\usepackage{placeins}
\usepackage{nicefrac}
\usepackage{braket}
\usepackage{pdfcomment}
\usepackage{xcolor}

\newcommand{\bk}{\mathbf{k}}

\newcommand{\angstrom}{\mbox{\normalfont\AA}}

\begin{document}

\title{Electric-field switchable second-harmonic generation in bilayer MoS$_2$ by inversion symmetry breaking}
%
\author{J.~Klein}\email{julian.klein@wsi.tum.de}
\affiliation{Walter Schottky Institut and Physik Department, Technische Universit\"at M\"unchen, Am Coulombwall 4, 85748 Garching, Germany}
\author{J.~Wierzbowski}
\affiliation{Walter Schottky Institut and Physik Department, Technische Universit\"at M\"unchen, Am Coulombwall 4, 85748 Garching, Germany}
\author{A.~Steinhoff}
\affiliation{Institut für Theoretische Physik, Universität Bremen, P.O. Box 330 440, 28334 Bremen, Germany}
\author{M.~Florian}
\affiliation{Institut für Theoretische Physik, Universität Bremen, P.O. Box 330 440, 28334 Bremen, Germany}
\author{M.~R\"osner}
\affiliation{Institut für Theoretische Physik, Universität Bremen, P.O. Box 330 440, 28334 Bremen, Germany}
\affiliation{Bremen Center for Computational Materials Science, Universität Bremen, 28334 Bremen, Germany}
\author{F.~Heimbach}
\affiliation{Lehrstuhl für Physik funktionaler Schichtsysteme, Physik Department E10, Technische Universit\"at München, James-Franck-Straße 1, 85748 Garching, Germany}
\author{K.~M\"uller}
\affiliation{Walter Schottky Institut and Physik Department, Technische Universit\"at M\"unchen, Am Coulombwall 4, 85748 Garching, Germany}
\author{F.~Jahnke}
\affiliation{Institut für Theoretische Physik, Universität Bremen, P.O. Box 330 440, 28334 Bremen, Germany}
\author{T.~O.~Wehling}
\affiliation{Institut für Theoretische Physik, Universität Bremen, P.O. Box 330 440, 28334 Bremen, Germany}
\affiliation{Bremen Center for Computational Materials Science, Universität Bremen, 28334 Bremen, Germany}
\author{J.~J.~Finley}\email{finley@wsi.tum.de}
\affiliation{Walter Schottky Institut and Physik Department, Technische Universit\"at M\"unchen, Am Coulombwall 4, 85748 Garching, Germany}
\author{M.~Kaniber}\email{kaniber@wsi.tum.de}
\affiliation{Walter Schottky Institut and Physik Department, Technische Universit\"at M\"unchen, Am Coulombwall 4, 85748 Garching, Germany}
%
%
\date{\today}
%
%

\begin{abstract}
\textbf{We demonstrate pronounced electric-field-induced second-harmonic generation in naturally inversion symmetric 2H stacked bilayer MoS$_{2}$ embedded into microcapacitor devices. By applying strong external electric field perturbations ($\left|F\right| = \pm \SI{2.6}{\mega\volt\per\centi\meter}$) perpendicular to the basal plane of the crystal we control the inversion symmetry breaking and, hereby, tune the nonlinear conversion efficiency. Strong tunability of the nonlinear response is observed throughout the energy range ($E_{\omega} \sim \SI{1.25}{\electronvolt} - \SI{1.47}{\electronvolt}$) probed by measuring the second-harmonic response at $E_{2\omega}$, spectrally detuned from both the A- and B-exciton resonances. A 60-fold enhancement of the second-order nonlinear signal is obtained for emission at $E_{2\omega} = \SI{2.49}{\electronvolt}$, energetically detuned by $\Delta E = E_{2\omega} - E_C = \SI{-0.26}{\electronvolt}$ from the C-resonance ($E_{C} = \SI{2.75}{\electronvolt}$). The pronounced spectral dependence of the electric-field-induced second-harmonic generation signal reflects the bandstructure and wave function admixture and exhibits particularly strong tunability below the C-resonance, in good agreement with Density Functional Theory calculations. Moreover, we show that the field-induced second-harmonic generation relies on the interlayer coupling in the bilayer. Our findings strongly suggest that the strong tunability of the electric-field-induced second-harmonic generation signal in bilayer transition metal dichalcogenides may find applications in miniaturized electrically switchable nonlinear devices.
}
\end{abstract}

%
%
\maketitle
%
%

Symmetry dictates the fundamental light-matter interactions in atoms, molecules and crystalline solids, including transition selection rules, optical activity and nonlinear optical response. Bulk crystals of 2H stacked transition metal dichalcogenides (TMDCs) are centrosymmetric giving rise to only a weak second-order nonlinear response due to a vanishing second-order electric susceptibility $\chi^{(2)}$ reflecting the inversion symmetric crystal structure.~\cite{Shen.1984,Wagoner.1998} However, in their few-layer forms these materials have a symmetry that is sensitive to the precise number of atomic layers.~\cite{Wilson.1969,Li.2013,Zeng.2013} Crystals with an even number of layers have macroscopic inversion symmetry whereas those with an odd number of layers lack a perfect inversion centre. Since TMDCs are van der Waals bonded two-dimensional (2D) materials, they can be mechanically exfoliated with either an even or odd number of layers.~\cite{Novoselov.2004} Bilayer crystals are naturally inversion symmetric systems in which local electric fields can be used to controllably break symmetries and alter the related optical and electrical properties. First experimental schemes for controlled symmetry breaking in 2D materials employing electric field effects have been demonstrated in bilayer graphene, allowing for band gap opening~\cite{Ohta.2006,Castro.2007,Zhou.2007,Zhang.2009}, while theoretical work suggests strongly tunable second-harmonic generation (SHG) in the mid-infrared by using in-plane or out-of-plane electric fields.~\cite{Wu.2012,Brun.2015} Likewise, symmetry breaking in bilayer MoS$_{2}$ for switching valley optical selection rules~\cite{Wu.2013} or valley Hall effects~\cite{Lee.2016} has been demonstrated. Moreover, very large second-order susceptibilities $\chi^{(2)}$  $\sim$ $\SI{0.1}{\nano\meter\per\volt}$ - $\SI{100}{\nano\meter\per\volt}$ for monolayer TMDCs have been reported due to the explicitly broken inversion symmetry; more than one order of magnitude larger than LiNbO$_{3}$. ~\cite{Li.2013,Malard.2013,Kumar.2013,Gruning.2014,Trolle.2014} Local structural defects that perturb crystal symmetry can also be probed using SHG since the signal is sensitive to local crystal orientation,~\cite{Li.2013,Malard.2013}  symmetry~\cite{Zeng.2013} or the presence of crystal edge states~\cite{Yin.2014}.

%
\begin{figure*}[!ht]
\scalebox{\figurescale}{\includegraphics[width=0.95\linewidth]{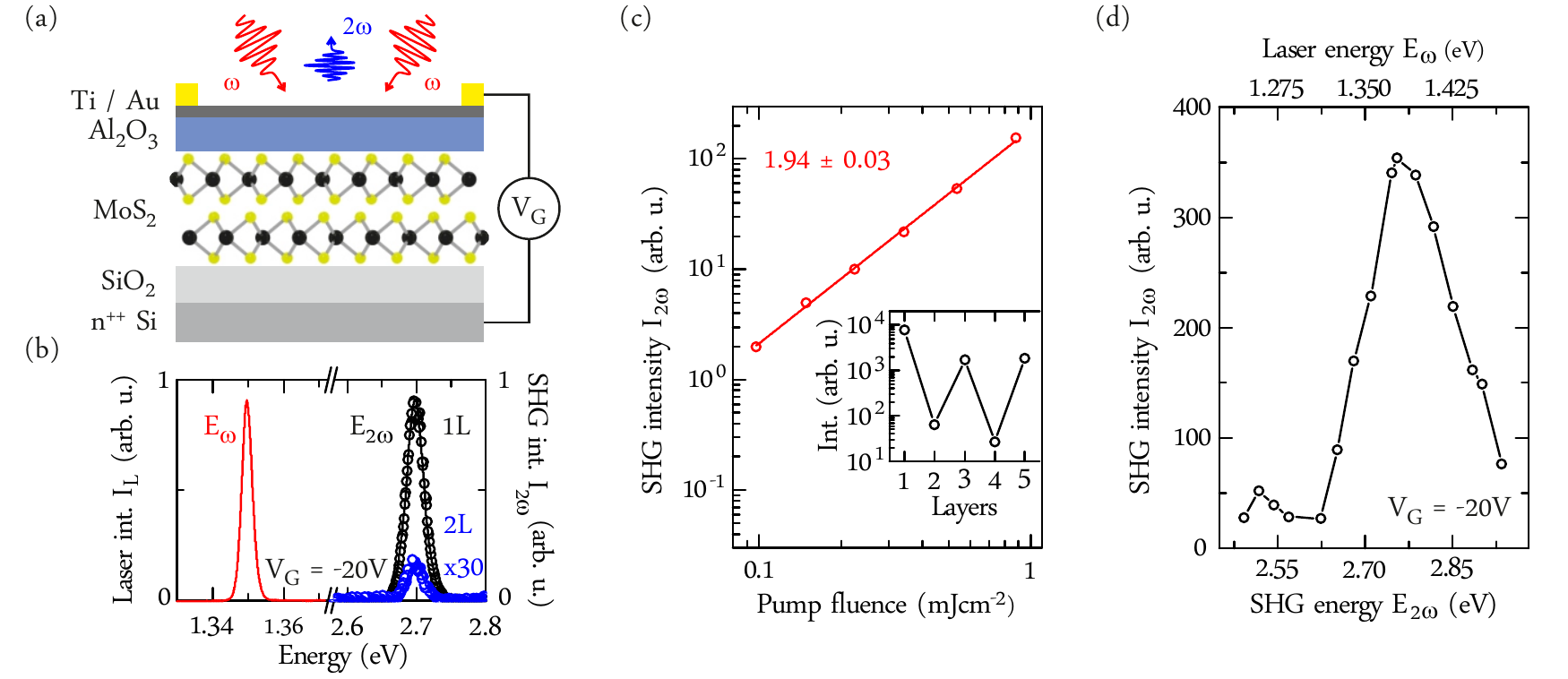}}
\renewcommand{\figurename}{Fig.}
\caption{\label{fig1}
(a) Schematic illustration of the bilayer MoS$_{2}$ microcapacitor device. The semi-transparent top gate facilitates optical measurements while applying a gate voltage $V_{G}$.
(b) Typical spectrum showing the excitation laser at $E_{\omega} = \SI{1.35}{\electronvolt}$ and the corresponding SHG signal at $E_{2\omega}$ for mono- (black circles) and bilayer (blue circles, 30 times magnified) 2H stacked MoS$_{2}$ at $V_{G} = \SI{-20}{\volt}$ where inversion symmetry is restored.
(c) Spectrally integrated SHG signal from a bilayer MoS$_{2}$ as a function of the laser pump fluence yielding a quadratic dependence. The inset shows the layer number dependent SHG intensity for mono- to pentalayer MoS$_{2}$ for a pump fluence of $\sim \SI{0.88}{\milli\joule\per\centi\meter\squared}$ (cw equivalent excitation power density of $\sim \SI{700}{\kilo\watt\per\centi\meter\squared}$).
(d) SHG dependence on the SHG emission energy $E_{2\omega}$ (laser excitation energy $E_\omega)$ at \SI{10}{\kelvin} at $V_G = \SI{-20}{\volt}$.
}
\end{figure*}
%

Recently, electrical control of the second-order nonlinear response has been reported in monolayers of WSe$_{2}$ using field effect devices. Hereby, tuning of the electron carrier density resulted in a $\sim$ 10-fold enhancement of the SHG signal when the second-harmonic emission energy $E_{2\omega}$ is on-resonance with the A-exciton.~\cite{,Seyler.2015} Similar device geometries were used for tuning the SHG response of a bilayer WSe$_{2}$ crystal by charge-induced second-harmonic generation (CHISH).~\cite{Yu.2015}  \\

In this letter, we report a $\sim$ 60$\times$ enhancement of the second-order nonlinear response of a natural inversion symmetric 2H stacked bilayer MoS$_{2}$ crystal by electric-field-induced second-harmonic generation (EFISH) in the vicinity of the C-resonance. The EFISH signal can be tuned using an external voltage applied to a microcapacitor device, which enables the application of large DC electric fields $F$ perpendicular to the basal plane of a bilayer MoS$_{2}$ crystal.~\cite{Klein.2016} We demonstrate continuous breaking and restoring of the crystal inversion symmetry for a wide range of fundamental laser excitation energies $E_{\omega} \sim \SI{1.25}{\electronvolt} - \SI{1.47}{\electronvolt}$. The voltage (field) dependence of the electrically induced SHG signal $I_{2\omega}$ from the bilayer TMDC is found to be quadratic, similar to observations in bulk crystals.~\cite{Terhune.1962,Bjorkholm.1967,Lee.1967} Moreover, calculations of the field-dependent second-order
susceptibility ($I_{2\omega} \propto \left|\chi^{(2)}\right|^2$) based on Density Functional Theory (DFT) are in good agreement with experimental observations, yielding the same quadratic applied-field dependence ($I_{2\omega} \propto\left|\chi^{(2)}\right|^2 \propto F^2$). We also find that hybridization of the two MoS$_{2}$ layers is a neccessary requirement to obtain an EFISH signal. Our work indicates that bilayer TMDCs are highly promising for electrically tunable nonlinear optical elements.\\

A schematic illustration of the device used in this letter is presented in Fig.~\ref{fig1}a. Mono- and few-layer crystals of 2H stacked MoS$_{2}$ are transferred onto $\SI{290}{\nano\meter}$ thick SiO$_{2}$ and capped with $\SI{20}{\nano\meter}$ of Al$_{2}$O$_{3}$ acting as top dielectric material.~\cite{Klein.2016} A $\SI{4.5}{\nano\meter}$ thick semi-transparent Ti window is lithographically defined on top of the Al$_{2}$O$_{3}$ and contacted with a Au ring to provide a homogeneous field distribution. The device facilitates application of a bias voltage V$_{G}$ between the top gate and the strongly n-doped Si substrate, leading to an electric field perpendicular to the basal plane of the embedded MoS$_{2}$ crystal. Recently, the same device geometry was used to demonstrate tunability of the A-exciton signal via the quantum confined Stark effect.~\cite{Klein.2016}

%
\begin{figure}[!ht]
\scalebox{\figurescale}{\includegraphics[width=0.95\linewidth]{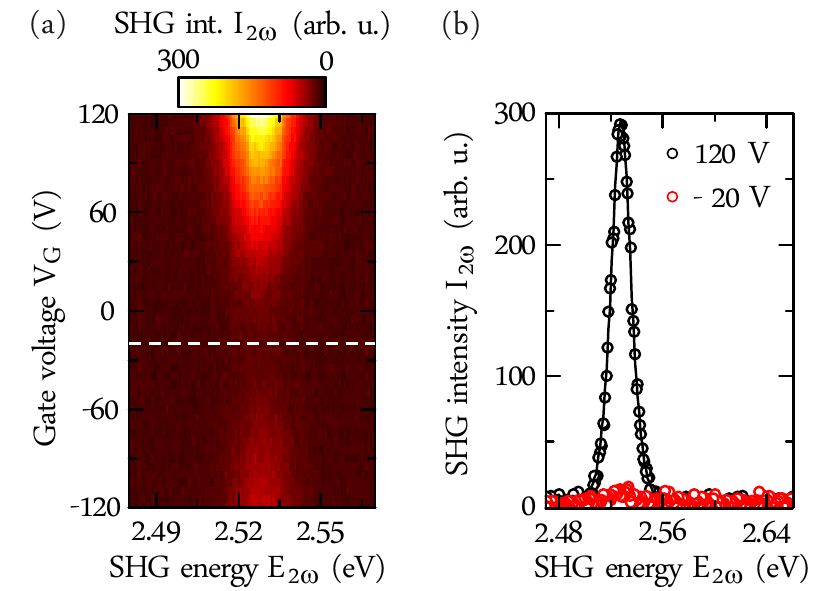}}
\renewcommand{\figurename}{Fig.}
\caption{\label{fig2}
(a) SHG intensity as a function of applied gate voltage and SHG emission energy of bilayer MoS$_{2}$ excited at $E_{\omega} = \SI{1.246}{\electronvolt}$. The dashed white line at $V_{G} = \SI{-20}{\volt}$ denotes the gate voltage where minimum SHG emission is observed.
(b) SHG spectra: Comparing the maximum ($V_{G} = \SI{120}{\volt}$, black) and minimum ($V_{G} = \SI{-20}{\volt}$, red) shows a clear voltage dependence of the SHG intensity. For $V_{G} = \SI{\pm 120}{\volt}$ the crystal inversion symmetry is maximally broken while inversion symmetry is restored at $V_{G} = \SI{-20}{\volt}$.
}
\end{figure}

A typical laser excitation spectrum at $E_{\omega} \sim \SI{1.35}{\electronvolt}$ and the corresponding SHG spectra of mono- and bilayer (30 times magnified) regions of the sample are shown in Fig.~\ref{fig1}b with fits using sech$^{2}$–functions. The spectra shown were recorded for a gate voltage $V_{G} = \SI{-20}{\volt}$ which corresponds to the point where inversion symmetry is maximally restored for the bilayer crystal embedded in the microcapacitor device as will be discussed in detail later. The monolayer region of the crystal exhibits a SHG response that is more than two orders of magnitude stronger than the bilayer region, an observation attributed to the explicitly broken inversion symmetry in the single layer limit.~\cite{Li.2013,Malard.2013,Kumar.2013,Janisch.2014} Since the bulk crystal used for our studies is 2H stacked, subsequent layers are rotated by $\SI{180}{\degree}$ with respect to each other.~\cite{Mattheiss.1973} As a direct consequence, the crystal space group alternates between even and odd number of layers:
TMDC crystals with an odd number of layers exhibit the D$_{3h}$ space group including both rotational and mirror symmetries~\cite{Shen.1984}, while crystals with an even layer number belong to D$_{3d}$ which additionally contains inversion symmetry leading to the strong difference in SHG signal as shown in Fig.~\ref{fig1}b.  Notably, a nonzero SHG signal in even layers is reported to arise from a phase difference between the two layers~\cite{Li.2013} or can further originate through crystal defects that locally break inversion symmetry. The power dependence of the SHG signal of inversion symmetric bilayer MoS$_{2}$ exhibits a clear quadratic dependence as shown by the red curve in Fig.~\ref{fig1}c. The SHG signal intensity of mono- to pentalayer regions of the MoS$_{2}$ flake for a constant pump fluence of $\sim \SI{0.88}{\milli\joule\per\centi\meter\squared}$ (cw equivalent of $\sim \SI{700}{\kilo\watt\per\centi\meter\squared}$), shown in the inset of Fig.~\ref{fig1}c, clearly oscillates with layer number,
confirming the strong dependence of the SHG signal on the crystal symmetry as expected. All following measurements within this letter were performed using a constant pump fluence of $\sim \SI{0.88}{\milli\joule\per\centi\meter\squared}$ which was chosen such that a finite SHG signal is always observed, independent of the gate voltage applied. The spectrally integrated SHG intensity recorded as a function of the SHG emission energy (excitation laser energy) for bilayer MoS$_{2}$ is presented in Fig.~\ref{fig1}d for a gate voltage of $V_{G} = \SI{-20}{\volt}$ that partially restores inversion symmetry. We observe a strong dependence of the SHG intensity on the laser energy, giving rise to a broad peak with a pronounced maximum at $E_{C} = \SI{2.75}{\electronvolt}$. The observed maximum spectrally coincides with the C-resonance, which can be qualitatively described both by excitonic effects~\cite{Qiu.2013,steinhoff_influence_2014} and by the band structure in a free-particle picture where it is associated with
a high joint density of states in the vicinity of the $\Gamma$ point in the Brillouin zone (BZ) as discussed in the Supporting Information.~\cite{Trolle.2014,Trolle.2015} The C-resonance is common in various TMDCs and is accompanied with a high optical absorption. The SHG signal at the C-resonance for bilayer MoS$_{2}$ is enhanced by approximately one order of magnitude as compared to off-resonant excitation, in good agreement with recent observations on noncentrosymmetric mono- and trilayer MoS$_{2}$.~\cite{Malard.2013} Moreover, the electronic structure and, therefore, the nonlinear optical response of the C-resonance is very robust with respect to changes in the dielectric environment as introduced by the Al$_{2}$O$_{3}$ capping layer (see pump-laser energy dependent SHG of capped and pristine monolayer MoS$_{2}$ in the Supporting Information). \\

In the following, we demonstrate electrical control of the SHG signal arising from the bilayer region of the sample by probing its intensity as a function of the applied voltage in the range - 120 V < V$_{G}$ < + 120 V. A false color plot of the SHG intensity of gated bilayer MoS$_{2}$ when subject to excitation at $E_{\omega} = \SI{1.246}{\electronvolt}$ is presented in Fig.~\ref{fig2}a. For later discussion, the excitation energy $E_{\omega}$ can also be translated into a detuning $\Delta E=E_{2\omega}-E_C=\SI{-0.26}{\electronvolt}$ of the corresponding second-harmonic energy with respect to the characteristic energy of the C-resonance $E_C$. Clearly, the SHG intensity for the maximum applied gate voltages $V_{G} = \SI{\pm 120}{\volt}$ is significantly enhanced as compared to zero applied bias at $V_{G} = \SI{0}{\volt}$. Notably, the minimum SHG intensity is observed at $V_{G} = \SI{-20}{\volt}$ which reflects the cancellation of the built-in field due to polarization charges in the surrounding dielectric
environment that is manifested by a field offset $F_0$. The offset has already been observed in DC Stark measurements of the A-exciton in mono- and few-
layer MoS$_{2}$ using the same device.~\cite{Klein.2016} The corresponding spectra for $V_{G} = \SI{+120}{\volt}$ (black circles) and $V_{G} = \SI{-20}{\volt}$ (red circles) are presented in Fig.~\ref{fig2}b. We attribute this characteristic tuning pattern to an electrically controlled breaking and restoration of the crystal inversion symmetry, mediated by the external voltage (electric field) applied to the bilayer MoS$_{2}$ crystal. This is in strong contrast to Refs. ~\cite{Seyler.2015} and ~\cite{Yu.2015} where the SHG tunability emerges from electrostatic charge doping dependent phenomena in TMDC flakes connected to a charge carrier reservoir by an electrical contact. In our experiment the TMDC flake is fully electrically isolated from the contacts due to the surrounding dielectric layers.  In this particular case, the net dipole moment of the bilayer is zero due to the oppositely oriented in-plane dipole moments of both layers. However, for a nonzero effective applied electric field $F_{\textrm{eff}} = F - F_0 \neq \SI{0}{\mega\volt\per\centi\meter}$, the crystal inversion symmetry is explicitly broken, analogous to the monolayer MoS$_{2}$. \\

%
\begin{figure}[!ht]
\scalebox{\figurescale}{\includegraphics[width=0.95\linewidth]{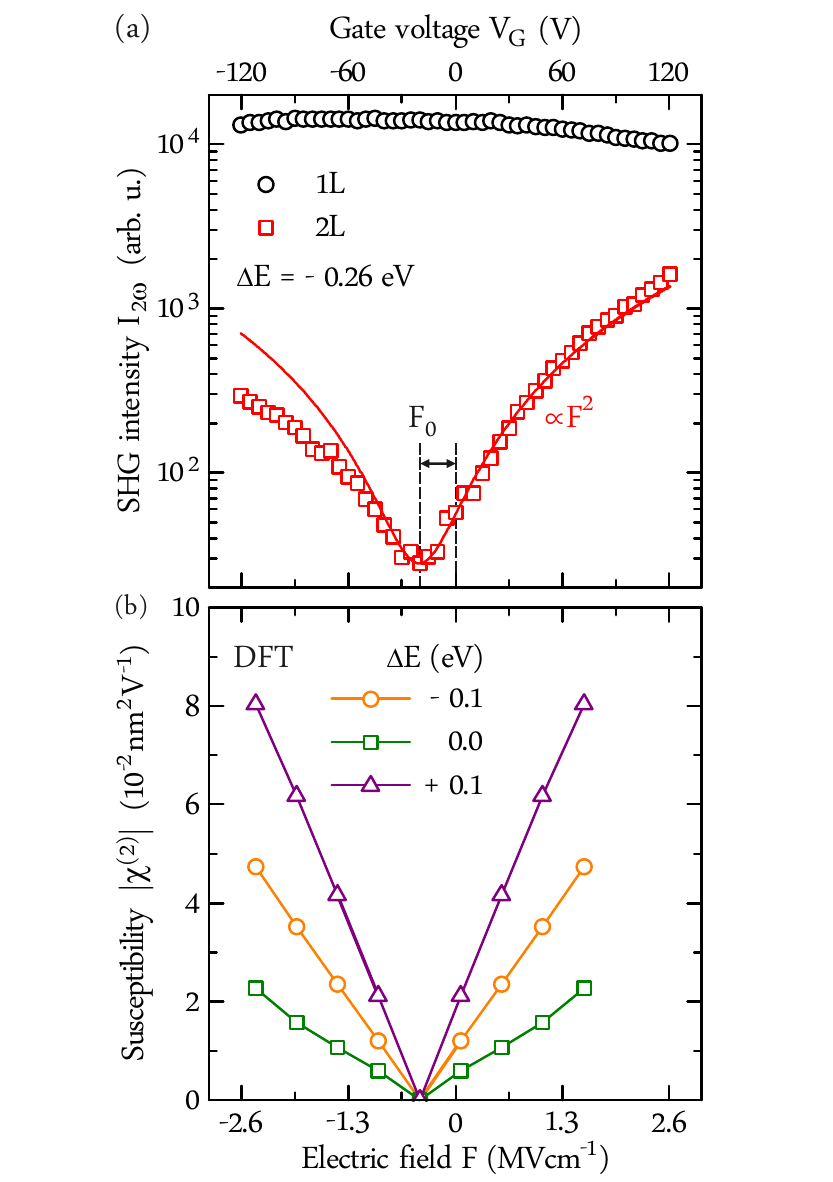}}
\renewcommand{\figurename}{Fig.}
\caption{\label{fig3}
(a) Measured electric field (gate voltage) dependent SHG intensity of mono- (black circles) and bilayer (red squares) MoS$_{2}$ at a detuning $\Delta E = E_{2\omega} - E_C = \SI{-0.26}{\electronvolt}$, which corresponds to the excitation shown in Fig.~\ref{fig2}. The field offset $F_0 = \SI{-0.44}{\mega\volt\per\centi\meter}$ ($V_{G} = \SI{-20}{\volt}$) is indicated by the dashed black lines where inversion symmetry is maximally restored. The red line is a quadratic fit to the data.
(b) Calculated $\left|\chi^{(2)}\right|$ (obtained from second-order susceptibility (Eq.~(\ref{eq:chi_2})) as described in the main text) for detunings of $\Delta E = \SI{0}{\electronvolt}$ (green squares), $\Delta E = \SI{-0.10}{\electronvolt}$ (orange circles)  and $\Delta E = \SI{+0.10}{\electronvolt}$ (purple triangles) as a function of external electric field strength. The theoretical data were shifted by $F_0$ for clarity.
}
\end{figure}

A detailed analysis of the gate voltage dependent SHG intensity of mono- (black circles) and bilayer (red squares) is presented in the Fig.~\ref{fig3}a on a semi-logarithmic scale. Here, we used 3D finite element (COMSOL) simulations to convert the applied gate voltage in electric field strengths.~\cite{Klein.2016} For the bilayer we observe a strong decrease of the SHG intensity when sweeping from large electric fields ($F = \SI{\pm 2.6}{\mega\volt\per\centi\meter}$) towards $F = \SI{0}{\mega\volt\per\centi\meter}$ with a pronounced minimum at $F_0 = \SI{-0.44}{\mega\volt\per\centi\meter}$ as indicated by the dashed black lines. The symmetry breaking in the bilayer leads to the observed strongly induced nonlinear response. In strong contrast, the monolayer exhibits a completely different characteristic field dependence. The lack of inversion symmetry manifests itself by an almost field independent SHG response owing to its symmetry point group. Similar data for a trilayer crystal is shown in the Supporting Information. For the bilayer, we observe an approximately 60-fold (11-fold) enhanced nonlinear response under nonresonant excitation at $E_{\omega} = \SI{1.246}{\electronvolt}$ at $F = \SI{+2.6}{\mega\volt\per\centi\meter}$ ($F = \SI{-2.6}{\mega\volt\per\centi\meter}$) compared to the inversion symmetric case at $F = \SI{-0.44}{\mega\volt\per\centi\meter}$ within the electric fields (voltages) applied in our experiments. For strongly broken inversion symmetry at $F = \SI{\pm 2.6}{\mega\volt\per\centi\meter}$ the SHG intensity almost reaches similar intensities as the monolayer. Since the data in Fig.~\ref{fig3} shows no sign of saturation, it indicates that one can obtain switchable SHG response from the bilayer producing a response comparable to a monolayer that lacks inversion symmetry. Moreover, the field (voltage) dependent SHG signal is in good agreement with an expected quadratic dependence $I_{2\omega} (V) \propto V^{2} \propto F^{2}$ as indicated by the fit in Fig.~\ref{fig3}a, which means that the susceptibility scales linearly with the applied voltage (electric field). This dependence was first reported in early work on EFISH~\cite{Terhune.1962,Bjorkholm.1967,Lee.1967} and later observed in molecules~\cite{Ward.1975}, solid state systems~\cite{Aktsipetrov.1994,Lupke.1995,Aktsipetrov.1996} and gases~\cite{Finn.1971,Bigio.1974}.

%
%
\begin{figure}[!ht]
\scalebox{\figurescale}{\includegraphics[width=0.95\linewidth]{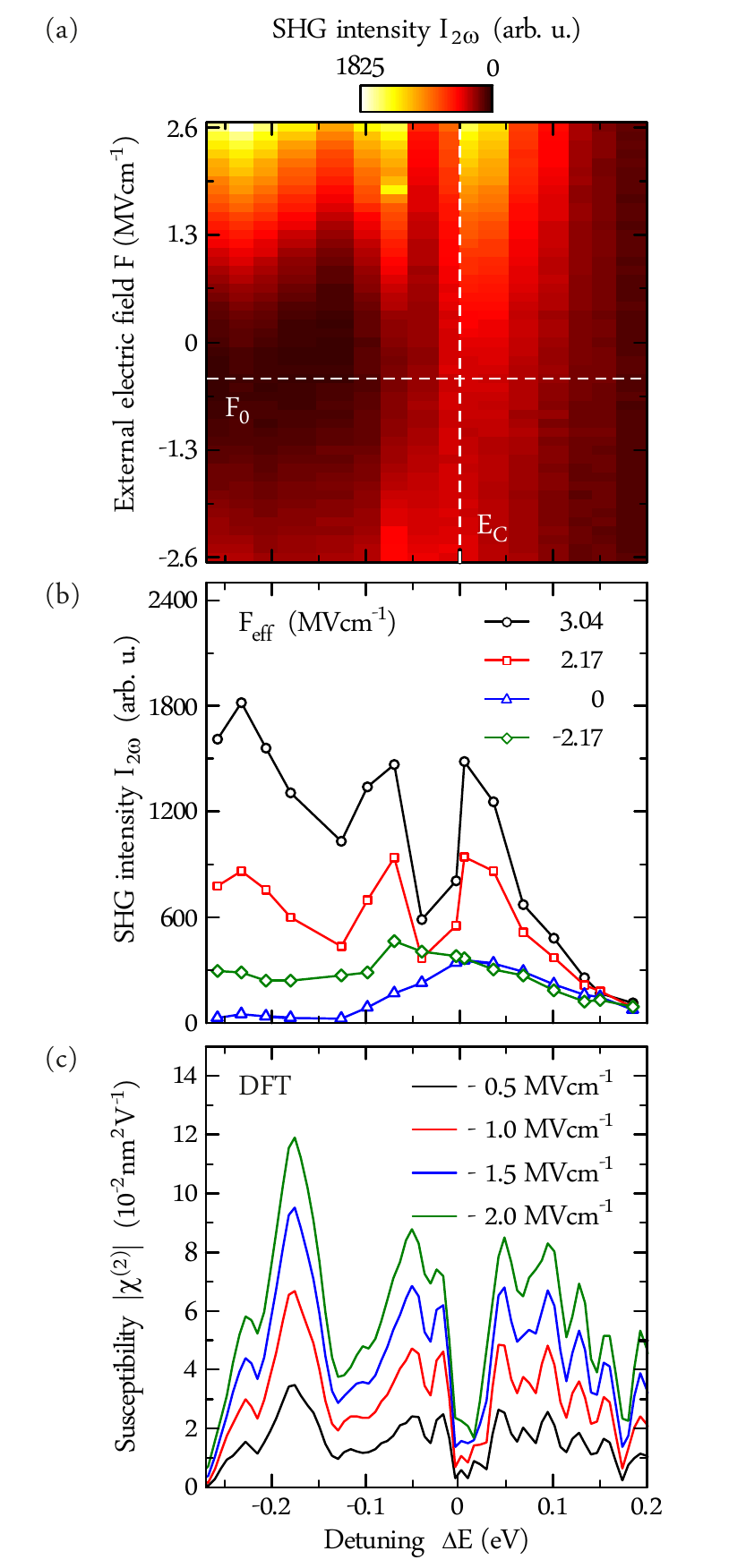}}
\renewcommand{\figurename}{Fig.}
\caption{\label{fig4}
(a) Integrated SHG intensity as a function of the external electric field and detuning $\Delta E = E_{2\omega} - E_C$. The field offset is at $F_{0} = \SI{-0.44}{\mega\volt\per\centi\meter}$ and the energy of the C-resonance $E_C = \SI{2.75}{\electronvolt}$ as identified from experiment are both highlighted with a dashed white line.
(b) SHG intensity as a function of detuning at effective electric fields $F_{\textrm{eff}} = \pm \SI{2.17}{\mega\volt\per\centi\meter}$ (red and green curve) and $F_{\textrm{eff}} = \SI{3.04}{\mega\volt\per\centi\meter}$ (black curve) with broken inversion symmetry and at $F_{\textrm{eff}} = \SI{0}{\mega\volt\per\centi\meter}$ (blue curve) with restored inversion symmetry.
(c) Calculated $\left|\chi^{(2)}\right|$ for different external electric field strengths as a function of detuning. The energy of the C-resonance is identified from linear response (see Supporting Information).
}
\end{figure}

To gain more insight into the origin of the electric field tunable SHG, we simulate the nonlinear response of bilayer MoS$_{2}$ under the influence of external electric fields, as described in the Methods section. Before considering nonlinear optical effects, however, we study the linear response of bilayer MoS$_2$ in the independent-particle approach (IPA) to identify the C-resonance $E_C$, which is discussed in more detail in the Supporting Information. This leads to a theoretical value of $E_C = \SI{2.57}{\electronvolt}$, which we use to express the second-harmonic emission energy $E_{2\omega}$ for the simulated results in terms of the detuning $\Delta E$, which has been introduced above for the experimental results. The calculated field dependent $\left|\chi^{(2)}\right|$ for detunings of $\Delta E = \SI{0}{\electronvolt}$, $\Delta E = \SI{-0.10}{\electronvolt}$ and $\Delta E = \SI{0.10}{\electronvolt}$ are shown in Fig.~\ref{fig3}b. As a general trend, a linear dependence on the external electric field
strength for fixed second-harmonic energy is observed, where the slope of the signal reflecting the tunability depends on the specific second-harmonic energy. This leads to a quadratic dependence of the measured SHG intensity on the applied field strength, as observed in the experimental data. Moreover, the magnitude of the $\left|\chi^{(2)}\right|$ is $\sim 5\times$ smaller than comparable theoretical results for monolayer MoS$_2$.~\cite{Trolle.2014} According to Eq.~(\ref{eq:P_2}), this means that we expect from theory a bilayer SHG intensity ($\propto P^2$) that is roughly an order of magnitude smaller than the monolayer signal if strong electric fields are applied. \\

In the final section of this letter, we investigate the electric field dependent SHG response of bilayer MoS$_2$ for a varying fundamental laser excitation energy. For this purpose, we swept the fundamental laser excitation energy $E_{\omega} \sim \SI{1.25}{\electronvolt} - \SI{1.47}{\electronvolt}$ in steps of $\sim \SI{14}{\milli\electronvolt}$ and measured the field dependent SHG response $E_{2\omega}$. Resulting SHG intensities as a function of external electric field and detuning with respect to the experimentally identified C-resonance at $E_C = \SI{2.75}{\electronvolt}$ are presented in a false color plot in Fig.~\ref{fig4}a. We observe a minimum signal intensity throughout all laser excitation energies at $F \sim \SI{-0.44}{\mega\volt\per\centi\meter}$ (corresponding to $V_G = \SI{-20}{\volt}$) as indicated by the white dashed horizontal line, in agreement with the built-in field offset discussed above. The maximum corresponding to the C-resonance at $E_{C} \sim \SI{2.75}{\electronvolt}$ for a fixed electric field is clearly visible in this color representation. Moreover, significant tunability of the SHG emission throughout the whole detected SHG signal energy range is observed. Cuts for constant effective applied electric fields of $F_{\textrm{eff}} = F - F_0 = \SI{3.04}{\mega\volt\per\centi\meter}$ (black), $F_{\textrm{eff}} = \SI{2.17}{\mega\volt\per\centi\meter}$ (red), $F_{\textrm{eff}} = \SI{-2.17}{\mega\volt\per\centi\meter}$ (green) and $F_{\textrm{eff}} = \SI{0}{\mega\volt\per\centi\meter}$ (blue) are shown in Fig.~\ref{fig4}b. For the highest electric fields applied, we observe a significant increase in SHG signal and additional finestructure in contrast to restored inversion symmetry at $F_{\textrm{eff}} = \SI{0}{\mega\volt\per\centi\meter}$ with only the C-resonance. Moreover, stronger tunability for negative detunings is observed while positive detunings yield a decreasing tunability. \\

%
%
\begin{figure}[!htb]
\scalebox{\figurescale}{\includegraphics[width=0.95\linewidth]{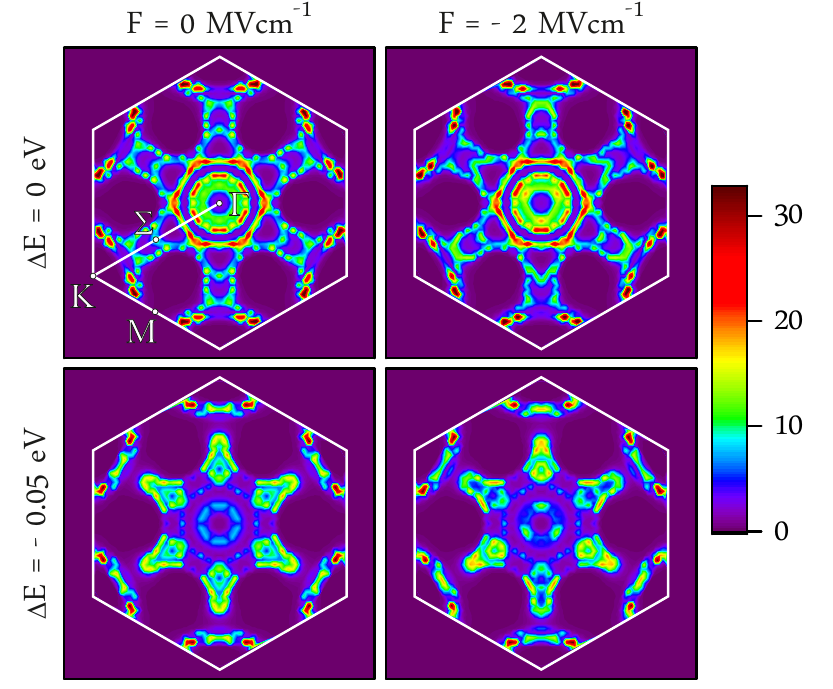}}
\renewcommand{\figurename}{Fig.}
\caption{\label{fig5}
Calculated wave functions without ($\SI{0}{\mega\volt\per\centi\meter}$) and with an external applied electric field ($\SI{-2}{\mega\volt\per\centi\meter}$) for detunings of $\Delta E = \SI{0}{\electronvolt}$ and $\Delta E = \SI{-0.05}{\electronvolt}$. Stronger symmetry breaking at $\Delta E = \SI{-0.05}{\electronvolt}$ leads to an enhanced tunability.
}
\end{figure}

We can qualitatively compare the experimental results with the corresponding calculated emission energy dependent susceptibility. Resulting susceptibilities from DFT simulations which correspond to the measured emission-energy dependent SHG spectra in Fig.~\ref{fig4}b, are shown in Fig.~\ref{fig4}c for $F = \SI{-0.5}{\mega\volt\per\centi\meter}$ (black), $F = \SI{-1}{\mega\volt\per\centi\meter}$ (red), $F =\SI{-1.5}{\mega\volt\per\centi\meter}$ (blue) and $F = \SI{-2}{\mega\volt\per\centi\meter}$ (green). The corresponding real and imaginary part of the susceptibility for $F = \SI{-0.5}{\mega\volt\per\centi\meter}$ and $F = \SI{-2}{\mega\volt\per\centi\meter}$ is shown in the Supporting Information. We find an energy dependence that is similar to the experimental spectra. The weak tunability around the C-resonance as compared to the tunability at $\Delta E = \SI{-0.05}{\electronvolt}$ can be understood from the different amount of symmetry breaking at the respective regions of the BZ where the SHG signal is generated. To visualize this effect, we calculate two-particle wave functions over the BZ using the procedure introduced in the Supporting Information and compare results with and without an
external electric field. In the absence of an electric field the wave functions exhibit the full sixfold symmetry expected from the crystal point group of the bilayer. By applying an electric field inversion symmetry is broken, as can be seen by comparing the left and right panels in Fig.~\ref{fig5}. At the C-resonance ($\Delta E = \SI{0}{\electronvolt}$, upper panels), we observe characteristic features between the $\Gamma$ and M points, which are only slightly modified under the influence of an external field. In turn, for a detuning of $\Delta E= \SI{-0.05}{\electronvolt}$ (lower panels) the wave functions show strong features around the $\Sigma$ point. Due to a different admixture of the atomic orbitals to the Bloch states in this part of the BZ the wave functions exhibit stronger symmetry breaking for non-vanishing electric fields in comparison to the C-resonance. This behaviour transfers to the stronger SHG tunability for emission energies below the C-resonance, which is observed in experiment.\\

We would like to point out that breaking the inversion symmetry of the bilayer e.g. by application of the external DC field is a neccessary but not sufficient condition to obtain a second-harmonic signal. An additional requirement is the hybridization of the two individual MoS$_2$ layers forming the bilayer that particularly mixes sulphur orbitals from both layers. As stated in Ref.~\cite{Yu.2015}, there is no out-of-plane third-order susceptibility tensor element for uncoupled TMDC layers in AB-stacking. Therefore, independent of an applied DC field, the nonlinear polarizations generated by two decoupled layers cancel due to their opposite orientation in the AB-stacking configuration. In the presence of hybridization though there is a formation of bonding and antibonding states that contain orbital character from both layers. If a DC field is applied in this situation, the nonlinear polarizations, which are microscopically electron-hole excitations, get a more hole-like character in the layer in direction of the DC field and vice versa. This leads to a net polarization from the bilayer and hence a nonzero SHG signal. We underline this effect by numerical calculation, comparing SHG spectra with applied field $F = \SI{-2}{\mega\volt\per\centi\meter}$ for coupled and decoupled MoS$_2$ layers (see interlayer spacing dependent calculation of $\chi^{(2)}$ in the Supporting Information). The decoupling of layers is realized by artificially doubling the interlayer distance in the DFT calculation and is verified by the fact that the band structure of two decoupled layers is given by two degenerate monolayer band structures (not shown). We find that the SHG signal of the decoupled layers is reduced by at least three orders of magnitude, which demonstrates the interplay between DC field and hybridization effects in the SHG from bilayer MoS$_2$.\\

In summary, we have demonstrated pronounced EFISH of the SHG intensity of 2H stacked bilayer MoS$_{2}$. Employing DFT we find a linear dependence of the $\left|\chi^{(2)}\right|$ (quadratic dependence of the SHG intensity) with applied electric field in good agreement with experiment. We find that hybridization between the two individual layers is a necessary requirement for generating EFISH. Our results show integrability of field effect devices combined with even-layered TMDCs that enable the electrical control and enhancement of nonlinear optical processes by electrically altering the crystal inversion symmetry. Potential schemes for prospective devices could exploit the switchable nonlinearities in bilayers for fast optical modulators.\\

\section{Methods}

\subsection{Second-harmonic measurements}

To probe the SHG signal arising from the mono-, and few-layer region of the MoS$_{2}$ flake investigated we excite the TMDC flake using a Ti:Sa oscillator that provides a pulse duration of $\sim \SI{120}{\femto\second}$, a repetition rate of $\sim \SI{80}{\mega\hertz}$ and a center frequency tunable over the energy range $E_{\omega} \sim \SI{1.25}{\electronvolt} - \SI{1.47}{\electronvolt}$, corresponding to a SHG signal energy of $E_{2\omega} \sim \SI{2.5}{\electronvolt} - \SI{2.94}{\electronvolt}$. The linearly polarised excitation laser is focused under normal incidence onto the sample surface with a microscope objective (NA = 0.50, 100x magnification) yielding a spot size of $d_{spot} \sim \SI{1.4}{\micro\meter}$ while the sample is held in a cryostat at $\SI{10}{\kelvin}$. The laser spot was placed in the very center of the flake to avoid probing of edge effects. The signal is detected along the same optical axis, without an analyser in the detection path, and dispersed onto a charge-coupled device (CCD) camera.

\subsection{Simulations}

The generation of the frequency-doubled optical polarization is described by the second-order susceptibility $\chi^{(2)}_{\alpha\beta\gamma}(-2\omega;\omega,\omega)$ according to
\begin{equation}
    P^{(2)}_{\alpha}(2\omega)=\varepsilon_0\chi^{(2)}_{\alpha\beta\gamma}(-2\omega;\omega,\omega)E_{\beta}(\omega)E_{\gamma}(\omega)~,
    \label{eq:P_2}
\end{equation}
where $\boldsymbol{E}(\omega)$ is the total macroscopic electric field and the subscripts denote cartesian coordinates. With the crystal in the x-y-plane and choosing the y-axis to point along an armchair direction of the TMDC bilayer, crystal symmetry requires that the only nonzero tensor components are $\chi^{(2)}\equiv \chi^{(2)}_{yyy}=-\chi^{(2)}_{xxy}=-\chi^{(2)}_{xyx}=-\chi^{(2)}_{yxx}$, as y- and z-axis span a mirror plane of the crystal. To calculate $\chi^{(2)}$, we use the formalism developed in ~\cite{leitsmann_second-harmonic_2005}, where second-order perturbation theory with respect to light-matter interaction and an independent-particle approach (IPA) is used to derive the expression
\begin{equation}
\begin{split}
&\chi^{(2)}_{\alpha\beta\gamma}(-2\omega;\omega,\omega)= \\
&\frac{-i e^3}{2\varepsilon_0(\omega+i\gamma)^3\hbar^2m^3A}\sum_{\bk}\sum_{nml}\frac{1}{\omega_{mn}(\bk)-2\omega-i\gamma} \\
&\times \Bigg[ \frac{f_{nl}(\bk)p^{\alpha}_{nm}(\bk)\left\lbrace p^{\beta}_{ml}(\bk)p^{\gamma}_{ln}(\bk)\right\rbrace}{\omega_{ln}(\bk)-\omega-i\gamma} \\
&+ \frac{f_{ml}(\bk)p^{\alpha}_{nm}(\bk)\left\lbrace p^{\gamma}_{ml}(\bk)p^{\beta}_{ln}(\bk)\right\rbrace}{\omega_{ml}(\bk)-\omega-i\gamma}\Bigg] ~.
\label{eq:chi_2}
\end{split}
\end{equation}
with
\begin{equation}
\left\lbrace p^{\beta}_{ml}(\bk)p^{\gamma}_{ln}(\bk)\right\rbrace=\frac{1}{2}\left[p^{\beta}_{ml}(\bk)p^{\gamma}_{ln}(\bk)+p^{\gamma}_{ml}(\bk)p^{\beta}_{ln}(\bk)\right]~.
\end{equation}
$\hbar\omega_{mn}(\bk)=\varepsilon_m(\bk)-\varepsilon_n(\bk)$ are the transition energies between band $m$ and $n$ at the point $\bk$ in the BZ, corresponding to the Bloch states $\ket{n\bk}$ and $\ket{m\bk}$, where $n$ includes the spin. $p^{\alpha}_{nm}(\bk)=\left<n\bk|\hat{p}^{\alpha}|m\bk\right>$ are momentum matrix elements and $f_{mn}(\bk)=f(\varepsilon_m(\bk))-f(\varepsilon_n(\bk))$ are Pauli blocking factors, which are assumed to be $0$ or $1$ for conduction and valence bands, respectively. The imaginary part of the frequency $\gamma$ accounts for a phenomenological broadening of resonances and $A$ is the crystal area.
Note that the susceptibility defined by Eq.~(\ref{eq:chi_2}) is a two-dimensional quantity, which means that the polarization induced in the material by the electric fields via Eq.~(\ref{eq:P_2}) is a dipole moment per unit area that is independent of the coordinate z. The polarization entering Maxwell's equations depending on z is therefore given by
\begin{equation}
    \boldsymbol{P}^{(2)}(z,2\omega)=\boldsymbol{P}^{(2)}(2\omega)\delta(z)~.
    \label{eq:P_2_maxwell}
\end{equation}
The $\delta$ function reflects the fact that the bilayer is assumed infinitely thin compared to the optical wavelength. By using Eq.~(\ref{eq:chi_2}), we assume that effects of a nonlocal potential~\cite{cabellos_effects_2009} that is used to obtain band structures and matrix elements are not essential to our results, which is reflected by the appearance of momentum instead of velocity matrix elements. In the same spirit, the momentum matrix elements can be
rewritten using the relation~\cite{gu_relation_2013}
\begin{equation}
\left<n\bk|\hat{\boldsymbol{p}}|m\bk\right>=im\omega_{nm}(\bk)\left<u_{n\bk}|i\nabla_{\bk}|u_{m\bk}\right>_{\textrm{cell}}~,
\end{equation}
where $\ket{u_{n\bk}}$ denote the lattice-periodic part of the Bloch states.

Assuming that quasi-particle effects, which can be described, e.g., on the level of a GW self-energy, and excitonic effects compensate to a large degree, we rely on a DFT based calculation to obtain the input data for the sum-over-states formula (\ref{eq:chi_2}). The band energies $\varepsilon_m(\bk)$ and matrix elements $\left<u_{n\bk}|i\nabla_{\bk}|u_{m\bk}\right>_{\textrm{cell}}$ are calculated within the PAW formalism~\cite{blochl_projector_1994} as described in Ref.~\cite{gajdos_linear_2006} and implemented in the Vienna Ab initio Simulation Package (VASP).~\cite{kresse_efficiency_1996, kresse_efficient_1996}  Therefore we apply the generalized gradient approximation ~\cite{perdew_generalized_1996} including spin-orbit coupling. The calculations are performed with AB-(Bernal)stacked bilayers of MoS$_2$ (lattice constant $a = \SI{3.18}{\angstrom}$, layer separation $c = \SI{6.35}{\angstrom}$, Mo-S z-separation $s_z \approx \SI{1.57}{\angstrom}$, super-cell height $h = \SI{20}{\angstrom}$) using 48x48x1 $\bk$-meshes, a plane-wave cutoff of $\SI{350}{\electronvolt}$, and a total amount of 60 bands. The electric field is applied perpendicular to the bilayer sheet and is varied between $0.000$ and $\SI{0.020}{\electronvolt\per\angstrom}$. In order to avoid interactions between the repeated slabs we apply dipole corrections to the potential as discussed in Ref.~\cite{neugebauer_adsorbate-substrate_1992}. As phenomenological broadening, a value of $\SI{20}{\per\pico\second}$ is used in Eq.~(\ref{eq:chi_2}) matching the experimental data well. We find that a total number of 28 valence bands and 16 conduction bands including spin are required to achieve convergence of the second-harmonic spectra.

%
%
\section{Acknowledgements}
This work has been supported by the Deutsche Forschungsgemeinschaft (DFG), in particular through the TUM International Graduate School of Science and Engineering (IGSSE). We gratefully acknowledge the BMBF for financial support by Q.com 16KIS0110. We gratefully acknowledge financial support of the German Excellence Initiative via the Nanosystems Initiative Munich and the PhD program ExQM of the Elite Network of Bavaria. We acknowledge resources for computational time at the HLRN (Hannover/Berlin) and support through the European Graphene Flagship.

\section{Author contributions}
J.K., J.W., M.K. and J.J.F. conceived and designed the experiments, J.K., J.W. and F.H. prepared the samples, J.K. performed the optical measurements, J.K. analyzed the data. A.S., M.F. and F.J. conceived and performed the SHG signal simulations based on input data from DFT calculations performed by M.R. and T.O.W. All authors contributed to the writing of the manuscript.

\section{Abbreviations}
TMDC, transition metal dichalcogenide; EFISH, electric-field induced second-harmonic generation; CHISH, charge-induced second-harmonic generation; 2D, two-dimensional; SHG, second-harmonic generation; DFT, Density Functional Theory; CCD, charge-coupled device; BZ, Brillouin zone;

\section{Supporting Information}
Pump-laser energy dependent SHG of capped and pristine monolayer MoS$_{2}$; voltage dependent SHG of mono- and trilayer MoS$_{2}$; determination of the energy of the C-resonance by theoretical evaluation of the linear susceptibility $\chi^{(1)}$; real and imaginary parts of $\chi^{(2)}$; interlayer spacing dependent $\chi^{(2)}$;

\section{Competing financial interests} The authors declare no competing financial interests.

%
%


\providecommand{\latin}[1]{#1}
\makeatletter
\providecommand{\doi}
  {\begingroup\let\do\@makeother\dospecials
  \catcode`\{=1 \catcode`\}=2\doi@aux}
\providecommand{\doi@aux}[1]{\endgroup\texttt{#1}}
\makeatother
\providecommand*\mcitethebibliography{\thebibliography}
\csname @ifundefined\endcsname{endmcitethebibliography}
  {\let\endmcitethebibliography\endthebibliography}{}

\end{document}